\documentclass[prc,floatfix,groupedaddress,nofootinbib,showpacs,preprintnumbers,
amsmath,amssymb,amsfonts,superscriptaddress,widetable] {revtex4}
\usepackage{graphicx}
\usepackage{dcolumn}
\usepackage{mathrsfs}
\usepackage{bm}

\usepackage{graphicx}
\usepackage{dcolumn}
\usepackage{bm}
\usepackage[usenames]{color}

\begin{document}

\title{Generic constraints on the Relativistic Mean-Field and Skyrme-Hartree-Fock models from the pure neutron matter equation of state}
\author{F. J. Fattoyev}
\email{Farrooh.Fattoyev@tamuc.edu} \affiliation{Department of
Physics and Astronomy, Texas A\&M University-Commerce, Commerce,
Texas 75429-3011, USA} \affiliation{Institute of Nuclear Physics,
Tashkent 100214, Uzbekistan}
\author{W. G. Newton}
\email{William.Newton@tamuc.edu} \affiliation{Department of Physics
and Astronomy, Texas A\&M University-Commerce, Commerce, Texas
75429-3011, USA}
\author{Jun Xu}
\email{Jun.Xu@tamuc.edu} \affiliation{Department of Physics and
Astronomy, Texas A\&M University-Commerce, Commerce, Texas
75429-3011, USA}
\author{Bao-An Li}
\email{Bao-An.Li@tamuc.edu} \affiliation{Department of Physics and
Astronomy, Texas A\&M University-Commerce, Commerce, Texas
75429-3011, USA}
\date{\today}
\begin{abstract}

We study the nuclear symmetry energy $S(\rho)$ and related
quantities of nuclear physics and nuclear astrophysics predicted
\emph{generically} by relativistic mean-field (RMF) and
Skyrme-Hartree-Fock (SHF) models. We establish a simple prescription
for preparing equivalent RMF and SHF parameterizations starting from
a minimal set of empirical constraints on symmetric nuclear matter,
nuclear binding energy and charge radii, enforcing equivalence of
their Lorenz effective masses, and then using the pure neutron
matter (PNM) equation of state (EoS) obtained from \emph{ab-initio}
calculations to optimize the pure isovector parameters in the RMF
and SHF models. We find the resulting RMF and SHF parameterizations
give broadly consistent predictions of the symmetry energy $J$ and
its slope parameter $L$ at saturation density within a tight range
of $\lesssim 2$ MeV and $\lesssim 6$ MeV respectively, but that
clear model dependence shows up in the predictions of higher-order
symmetry energy parameters, leading to important differences in (a)
the slope of the correlation between $J$ and $L$ from the confidence
ellipse, (b) the isospin-dependent part of the incompressibility of
nuclear matter $K_{\tau}$, (c) the symmetry energy at
supra-saturation densities, and (d) the predicted neutron star
radii. The model dependence can lead to about 1-2 km difference in
predictions of the neutron star radius given \emph{identical}
predicted values of $J$, $L$ and symmetric nuclear matter (SNM)
saturation properties. Allowing the full freedom in the effective
masses in both models leads to constraints of $30 \lesssim J
\lesssim 31.5$ MeV, $35 \lesssim L \lesssim 60$ MeV, $-330 \lesssim
K_{\tau} \lesssim -216$ MeV for the RMF model as a whole and $30
\lesssim J \lesssim 33$ MeV, $28 \lesssim L \lesssim 65$ MeV, $-420
\lesssim K_{\tau} \lesssim -325$ MeV for the SHF model as a whole.
Notably, given PNM constraints, these results place RMF and SHF
models \emph{as a whole} at odds with some constraints on $K_{\tau}$
inferred from giant monopole resonance and neutron skin experimental
results.

\end{abstract}
 \maketitle

\section{Introduction}
\label{Introduction}

Highly isospin asymmetric nuclear matter is present in heavy nuclei
far from the stability line and in the surface regions of nuclei
exhibiting neutron skins and occurs during heavy-ion collisions and
also in astrophysical systems such as neutron stars. Its energy
relative to that of symmetric nuclear matter (SNM) is usefully
characterized by the symmetry energy as a function of density
$S(\rho)$, and therefore the constraining of $S(\rho)$, particularly
at densities away from nuclear saturation density $\rho_0$, has been
the subject of much recent experimental and theoretical activity
\cite{Chen:2004si,Famiano:2006rb,Shetty:2007zg,Klimkiewicz:2007zz,
Danielewicz:2008cm,Tsang:2008fd,Centelles:2008vu,Warda:2009tc,
Carbone:2010az,Chen:2010qx,Zenihiro:2010zz,Xu:2010fh,
Liu:2010ne,Chen:2011ek,PhysRevLett.108.052501,Lattimer:2012xj,
Dong:2012zz,Piekarewicz:2012pp,Tsang:2012se}. Particularly,
constraints extracted for the magnitude of the symmetry energy at
saturation density $J \equiv S(\rho_0)$ and its slope there $L$ span
the ranges $J \approx  25-35$ MeV and $L \approx  25-115$ MeV
although the $J$ constraints from mass models alone are much
tighter, and most recent $L$ constraints are placed in the range
$30-80$ MeV (see, e.g., \cite{Tsang:2012se} and  Fig. 1 from
\cite{Li:2011xra}). Extraction of such constraints requires
specifying a model for nucleon-nucleon interactions which tends to
be equivalent to specifying a choice of functional form for
$S(\rho)$ in the RMF and SHF models. It is the influence of such
choices on predictions of symmetry energy and simple, related,
nuclear and astrophysical properties that will be the subject of
this paper.

Given the currently prohibitive complexity of computing the
nucleon-nucleon (NN) interaction from the underlying theory of QCD
and the challenges of solving the nuclear many-body problems,
progress in nuclear many-body theory has developed along two lines:

(1) The microscopic approach builds up the many body system from
bare NN interactions together with a usually phenomenological
description of the three-nucleon (3N) interaction; in-medium
correlations are self-consistently included via the many-body
calculational technique. In the last few years, much theoretical
progress in understanding neutron-rich systems has been accomplished
through microscopic pure neutron matter (PNM) calculations. By
studying the universal behavior of resonant Fermi gases with
infinite scattering length, a significant constraint is achieved for
the equation of state of dilute neutron
matter~\cite{Schwenk:2005ka}. These calculations have been extended
to higher densities using the full power of Quantum Monte Carlo
methods~\cite{Gezerlis:2009iw, Gezerlis:2011ai}. By studying the
physics of chiral three-nucleon forces the EoS of PNM is obtained
perturbatively up to nuclear saturation
density~\cite{Hebeler:2009iv}. Finally, the Auxiliary Field
Diffusion Monte Carlo (AFDMC) technique, which takes into account
the realistic nuclear Hamiltonian containing modern two- and
three-body interactions of the Argonne potential and Urbana family
of three-body nucleon forces, has  been used to calculate the EoS of
PNM up to and above saturation density~\cite{Gandolfi:2009fj,
Gandolfi:2009nq, Gandolfi:2011xu}. The outcome of the above
investigations is a robust prediction for the EoS of PNM at low
densities where only two body interactions are important, and a
systematic investigation of the uncertainties in the EoS of PNM up
to and beyond saturation density as a result of our present
uncertainties in the three-neutron interaction, resulting in
``theoretical error bars'' in that density regime.

(2) The second approach is to construct an effective interaction
describing the \emph{in-medium} nucleon-nucleon interaction, subject
to most of the symmetries of the bare potential. The effective
interaction is typically dependent on $\sim 10$ parameters
representing, for example, coupling constants, which are fit to
experimental data sets from finite nuclei properties such as binding
energies, charge radii, single particle energy spectra and spectra
of collective excitations. One of the overriding goals of modern
nuclear many-body theory is to derive an energy-density functional
(EDF)~\cite{UNEDF} with clear physical connections to \emph{ab
initio} NN interactions and QCD. The widely used
RMF~\cite{Serot:1984ey, Serot:1997xg} and SHF~\cite{Skyrme:1956zz,
Vautherin:1971aw} models, with the latter thought of as a
non-relativistic expansion of the former~\cite{Reinhard:1989zi,
Sulaksono:2003re}, are two typical phenomenological EDFs used in
nuclear many-body theory. Both models have $\lesssim 10$ free
parameters in their simplest forms. Recent surveys find about $240$
parameter sets for the SHF model~\cite{Dutra:2012mb} and 10s of
parameterizations of the simplest form of the RMF model,
e.g.~\cite{Chen:2007ih}, although many are old parameter sets
superseded by parameter sets fit to more recent, accurate data. We
shall refer to the space inhabited by the free parameters as the
model space, and the two classes of EDF (RMF and SHF) shall be
referred to as the two models. Since the number of experimental
observables is always larger than the number of free parameters, the
problem of optimizing these EDFs is always overdetermined, and this
results in a considerable degeneracy amongst parameter sets, and
correlations between individual parameters when constrained by
certain observables. Covariance analysis
techniques~\cite{Reinhard:2010wz, Fattoyev:2011ns} have been
employed to study correlations between predicted observables from a
particular EDF in its model space. Given a certain experimental
constraint, this analysis method serves as a useful tool to optimize
the parameters of RMF and SHF EDFs and systematically examine the
correlations between various nuclear matter and neutron star
properties \cite{Lattimer:2012xj}.

Most parameterizations of EDFs are obtained through fitting to
predominantly nuclear experimental data sets, thus probing closely
isospin symmetric systems; their predictions of symmetry energy
behaviors thus vary widely, and many do not give PNM predictions
consistent with microscopic calculations. Recent parameterizations
which do take into account \emph{ab-initio} PNM calculations tend to
give behaviors of the symmetry energy at odds with some experimental
constraints from giant monopole resonances and neutron skins
\cite{Centelles:2008vu,Pearson:2010zz,Garg:2011yr}; constraints
extracted, in part, using the same types of EDFs. This discrepancy
also occurs in microscopic calculations \cite{Vidana:2009is}. This
raises questions such as: can such discrepancies be resolved by
choosing a different parameterization? Are the extracted constraints
correct? Is there a fundamental problem with the particular EDFs
used? A systematic study of \emph{parameterization-independent} RMF
and SHF model predictions of the behavior of $J$ and $L$ and related
physical properties has yet to be undertaken.

RMF and SHF models predict different functional forms for $S(\rho)$
and thus one might expect generic differences in the values
extracted from the same sets of data within each model, and
conversely, generic differences in the predictions of properties of
neutron-rich systems, even given the same values of $J$ and $L$.
This potentially makes combining the two types of functional in
modeling experimental phenomena hazardous; it also means that
extraction of, for example, constraints on $J$ and $L$ from nuclear
experiment or astrophysical observation (e.g. the measurement of
neutron star radii) comes with the caveat that such constraints are
dependent on the model used for extraction. It is important to
attempt to quantify what difference that model choice makes.

The aim of this paper is to explore the generic predictions of
properties of isospin-asymmetric nuclear matter from RMF and SHF
EDFs simultaneously constrained by the best theoretical knowledge of
the PNM EoS. Particularly, we will set model-generic best fit values
and $1\sigma$ confidence intervals on $J$, $L$, $K_{\tau}$ and
neutron skin thicknesses arising from the optimization to the
theoretical PNM EoS; we will examine whether the discrepancies
between the predicted values of $K_{\tau}$ and those extracted from
experiment are endemic to the two EDFs as a whole, and we will
explore the model dependence of the above results, the
supra-saturation symmetry energy and neutron star properties arising
from the different functional forms of the two models. We should
note here that our aim is neither to establish new parameterizations
of either model, nor to set absolute constraints on symmetry energy,
but to explore as far as possible the \emph{generic} constraints
that can be placed by each model on neutron-rich systems once
constrained by information from the PNM EoS.

The manuscript is organized as follows. In Sec.~\ref{Formalism} we
briefly review the two EDFs and the covariance analysis method used
to optimize the two pure isovector parameters in the EDFs. Results
are presented in Sec.~\ref{Results} and in Sec.~\ref{summary} we
conclude.

\section{Formalism}
\label{Formalism}

To ease discussions, we recall in this section the main formulas
related to the symmetry energy, the RMF and SHF models, and the
covariance analysis method used to examine the effect of the PNM
constraints on the two models.

\subsection{Symmetry energy}

The binding energy per nucleon in neutron-rich nuclear matter can be
written as
\begin{equation}\label{eanm}
E(\rho,\alpha) = E_0(\rho) + S(\rho) \alpha^2 +
\mathcal{O}(\alpha^4),
\end{equation}
where $\rho$ is the baryon number density and $\alpha=(\rho_{\rm
n}-\rho_{\rm p})/\rho$ is the isospin asymmetry, with $\rho_{\rm n}
(\rho_{\rm p})$ being the neutron (proton) number density. Around
the saturation density $\rho_0$, the symmetry energy can be
expressed as
\begin{equation}
S(\rho) = J + L \chi + \frac{1}{2}K_{\rm sym}\chi^2 +
\mathcal{O}(\chi^3) \ ,
\end{equation}
where $\chi \equiv \left(\rho - \rho_0\right)/3\rho_0$, $J$ is the
value of the symmetry energy at saturation density, $L$ is the slope
parameter, and $K_{\rm sym}$ is the curvature parameter at
saturation density given, respectively, by the following
expressions:
\begin{eqnarray}
&&  L = 3 \rho_0 \left(\frac{\partial S(\rho)}{\partial
\rho}\right)_{\rho=\rho_0} \ , \\
&& K_{\rm sym} = 9 \rho_0^2 \left(\frac{\partial^2 S(\rho)}{\partial
\rho^2}\right)_{\rho=\rho_0} \ .
\end{eqnarray}
The coefficients of the higher-order terms in Eq.~(\ref{eanm}) are
generally much smaller than $S(\rho)$, so it is usually a good
approximation to write the energy per nucleon in PNM as $E_{\rm
PNM}(\rho) \approx E_0(\rho)+S(\rho)$; however, in this work we calculate $E_{\rm
PNM}$ using the full EoS.

\subsection{Relativistic mean-field model}

The commonly employed RMF model contains an isodoublet nucleon field
($\psi$) interacting via the exchange of the scalar-isoscalar
$\sigma$-meson ($\phi$), the vector-isoscalar $\omega$-meson
($V^{\mu}$), the vector-isovector $\rho$-meson (${\bf b}^{\mu}$),
and the photon ($A^{\mu}$)~\cite{Serot:1984ey, Serot:1997xg,
Mueller:1996pm}. The effective Lagrangian density for the model can
be written as
\begin{eqnarray}
{\mathscr L} &=& \bar\psi \left[\gamma^{\mu} \left(i \partial_{\mu}
\!-\! g_{\rm v}V_\mu  \!-\! \frac{g_{\rho}}{2}{\mbox{\boldmath
$\tau$}}\cdot{\bf b}_{\mu} \!-\!
\frac{e}{2}(1\!+\!\tau_{3})A_{\mu}\right)  \!-\! \left(M \!-\!
g_{\rm s}\phi\right) \right]\psi +
\frac{1}{2}\partial_{\mu}\phi\,\partial^{\mu} \phi
-\frac{1}{2}m_{\rm s}^{2}\phi^{2} \nonumber \\
&-& \frac{1}{4}V^{\mu\nu}V_{\mu\nu} + \frac{1}{2}m_{\rm
v}^{2}V^{\mu}V_{\mu} - \frac{1}{4}{\bf b}^{\mu\nu}\cdot{\bf
b}_{\mu\nu} + \frac{1}{2}m_{\rho}^{2}\,{\bf b}^{\mu}\cdot{\bf
b}_{\mu} -\frac{1}{4}F^{\mu\nu}F_{\mu\nu} -
          U(\phi,V_{\mu},{\bf b_{\mu}}) \;,
 \label{LDensity}
\end{eqnarray}
where $V_{\mu\nu}$, ${\bf b}_{\mu\nu}$, and $F_{\mu\nu}$ are the
isoscalar, isovector, and electromagnetic field tensors,
respectively:
\begin{subequations}
\begin{align}
 V_{\mu\nu} &= \partial_{\mu}V_{\nu} - \partial_{\nu}V_{\mu} \;, \\
 {\bf b}_{\mu\nu} &= \partial_{\mu}{\bf b}_{\nu}
 - \partial_{\nu}{\bf b}_{\mu} \;, \\
F_{\mu\nu} &= \partial_{\mu}A_{\nu} - \partial_{\nu}A_{\mu} \;.
\label{FieldTensors}
\end{align}
\end{subequations}
The nucleon mass $M$ and meson masses $m_{\rm s}$, $m_{\rm v}$, and
$m_{\rho}$ may be treated as empirical parameters. The effective
potential $U(\phi,V_{\mu},{\bf b_{\mu}})$ consists of non-linear
meson interactions that simulates the complicated dynamics encoded
in just few model parameters. In the present work we use the
following form of the effective potential~\cite{Todd-Rutel:2005fa}:
\begin{align}
  U(\phi,V^{\mu},{\bf b}^{\mu})  =
    \frac{\kappa}{3!} (g_{\rm s}\phi)^3 \!+\!
    \frac{\lambda}{4!}(g_{\rm s}\phi)^4 \!-\!
    \frac{\zeta}{4!}   g_{\rm v}^4(V_{\mu}V^\mu)^2 -
   \Lambda_{\rm v} g_{\rho}^{2}\,{\bf b}_{\mu}\cdot{\bf b}^{\mu}
           g_{\rm v}^{2}V_{\nu}V^\nu\;.
\label{USelf}
\end{align}
This model is described by $7$ interaction parameters: $\{g_{\rm s},
g_{\rm v}, g_{\rho}, \kappa, \lambda, \zeta, \Lambda_{\rm v}\}$.
Note that power counting suggests that a consistent Lagrangian
density should include all terms up to fourth order in the meson
fields. However, the existing database of both laboratory and
observational data appears to be accurately described by the the
minimal set of parameters~\cite{Lalazissis:1996rd, Lalazissis:1999,
Todd-Rutel:2005fa}. Indeed, it was shown that ignoring a subset of
model parameters that are of the same order in a power-counting
scheme does not compromise the quality of the
fit~\cite{Mueller:1996pm, Furnstahl:1996wv}.

\subsection{Skyrme-Hartree-Fock model}

The standard form of the energy density obtained from the zero-range
Skyrme interaction using the Hartree-Fock method can be written as~\cite{Chabanat:1997qh}
\begin{eqnarray}
{\mathscr H} &=& \frac{\hbar^2}{2M} \tau  \!+\!
t_0\left[\left(2+x_0\right)\rho^2 -
\left(2x_0+1\right)\left(\rho_{\rm n}^2 + \rho_{\rm
p}^2\right)\right]/4 \nonumber \\ &+&
t_3\rho^{\sigma}\left[\left(2+x_3\right)\rho^2 -
\left(2x_3+1\right)\left(\rho_{\rm n}^2 + \rho_{\rm
p}^2\right)\right]/24 \nonumber \\ &+& \left[t_2\left(2x_2+1\right)
-t_1\left(2x_1+1\right)\right]\left(\tau_n\rho_n +
\tau_p\rho_p\right)/8 + \left[t_1\left(2+x_1\right)
+t_2\left(2+x_2\right)\right]\tau\rho/8 \nonumber \\ &+&
\left[3t_1\left(2+x_1\right)
-t_2\left(2+x_2\right)\right]\left(\nabla \rho\right)^2/32 -
\left[3t_1\left(2x_1+1\right)+t_2\left(2x_2+1\right)\right]\left[\left(\nabla
\rho_{\rm n}\right)^2 + \left(\nabla \rho_{\rm p}\right)^2\right]/32
\nonumber \\ &+& W_0\left[\vec{J} \cdot \nabla\rho  + \vec{J}_{\rm
n} \cdot \nabla\rho_{\rm n} + \vec{J}_{\rm p} \cdot \nabla\rho_{\rm
p}\right]/2 + \left(t_1-t_2\right)\left[J_{\rm n}^2 + J_{\rm
p}^2\right]/16 -\left(t_1x_1 + t_2x_2\right)J^2/16 \ .
\end{eqnarray}
Here $\rho_q$, $\tau_q$, and $\vec{J}_q$ ($q={\rm p}, {\rm n}$) are,
respectively, the number, kinetic, and spin-current densities, and
$\rho$, $\tau,$ and $\vec{J}$ are the corresponding total densities.
The SHF model is expressed in terms of $9$ Skyrme parameters:
$\{t_0, t_1, t_2, t_3, x_0, x_1, x_2, x_3, \sigma\}$ and the
spin-orbit coupling constant $W_0$ which is taken as $133.3$ ${\rm
MeV} \, {\rm fm}^5$~\cite{Chen:2010qx} in the present work.

\subsection{Covariance analysis method}

Here we very briefly discuss the covariance analysis method used in
the present work. For more details, we refer the readers to
Refs.~\cite{Reinhard:2010wz, Fattoyev:2011ns, Brandt:1999}.
Given a set of $N$ experimental observables $\mathcal{O}_{n}^{\rm
(exp)}$ that are determined with an accuracy of
$\Delta\mathcal{O}_{n}$, one can minimize the quality measure
$\chi^2$:
\begin{equation}
 \chi^{2}({\bf p}) \equiv \sum_{n=1}^{N}
 \left(\frac{\mathcal{O}_{n}^{\rm (th)}({\bf p})-
 \mathcal{O}_{n}^{\rm (exp)}}
 {\Delta\mathcal{O}_{n}}\right)^{2} \;.
 \label{ChiSquare}
\end{equation}
Here each of the $N$ observables is computed within the given model
$\mathcal{O}_{n}^{\rm (th)}({\bf p})$ as a function of the $F$ model
parameters ${\bf p}\!=\!(p_{1},\ldots,p_{F})$. A set of optimal
parameters ${\bf p}_0$ are determined via a least square fit to the
quality measure. For our set of `experimental' observables
$\mathcal{O}_{n}^{\rm (exp)}$ in the $\chi^2$ input we choose the
theoretical calculations of the energy per neutron $E_{\rm PNM}$ in
the density range of $0.04 \leq \rho \leq 0.16$
fm$^{-3}$~\cite{Akmal:1998cf, Hebeler:2009iv, Gandolfi:2009fj}.
Although the AFDMC calculations have been extended up to several
times the saturation density~\cite{Gandolfi:2011xu}, the extension
of the calculations of the chiral three-nucleon forces to higher
densities using piecewise polytropes~\cite{Hebeler:2010jx} shows
that the uncertainties in the EoS could be very large when all of
these models are employed. Therefore we rely on the PNM calculations
that are obtained up to saturation density only. Moreover, the
symmetry energy coefficients are only sensitive to the equation of
state around the saturation density.

Once the optimal parameter set ${\bf p}_{0}$ is found through the
$\chi^{2}$-minimization, one can then compute and diagonalize the
symmetric matrix of second derivatives. All the information about
the behavior of the $\chi^2$ function around the minimum is
contained in this matrix. That is,
\begin{equation}
 \chi^2({\bf p}) -\chi^{2}({\bf p}_{0})
 \equiv \Delta\chi^2({\bf x}) =
 {\bf x}^{T}{\hat{\mathcal M}}\,{\bf x} = {\bm\xi}^{T}
   {\hat{\mathcal D}}{\bm\xi}=\sum_{i=1}^{F}
   \lambda_{i}\xi_{i}^{2} \;,
 \label{Taylor2}
\end{equation}
where
\begin{equation}
  x_{i} \equiv \frac{({\bf p}-{\bf p}_{0})_{i}}{({\bf p}_{0})_{i}} \;
 \label{xDef}
\end{equation}
are scaled dimensionless variables, ${\hat{\mathcal M}} =
{\hat{\mathcal A}}{\hat{\mathcal D}} {\hat{\mathcal A}^{T}}$, and
${\bm\xi}\!=\!{\hat{\mathcal A}^{T}}{\bf x}$ are dimensionless
variables in a rotated basis. Here ${\hat{\mathcal A}}$ is the
orthogonal matrix whose columns are composed of the normalized
eigenvectors and ${\hat{\mathcal D}}\!=\!{\rm diag}
(\lambda_{1},\ldots,\lambda_{F})$ is the diagonal matrix of
eigenvalues. The meaningful theoretical uncertainties can be
obtained by computing the statistical covariance of two observables
$A$ and $B$ which can be written as follows:
\begin{equation}
 {\rm cov}(A,B) = \sum_{i,j=1}^{F}
 \frac{\partial A}{\partial x_{i}}
  (\hat{{\mathcal M}}^{-1})_{ij}
 \frac{\partial B}{\partial x_{j}} =
 \sum_{i=1}^{F}
 \frac{\partial A}{\partial \xi_{i}}
 \lambda_{i}^{-1}
 \frac{\partial B}{\partial \xi_{i}} \;.
 \label{Covariance}
\end{equation}
The variance $\sigma^{2}(A)$ of a given observable $A$ is then
simply given by $\sigma^2(A)\!=\!{\rm cov}(A,A)$. Finally, the
covariance ellipses between two observables $A$ and $B$ can be
plotted by diagonalizing the $2 \times 2$ covariance matrix:
\begin{equation}
\hat{{\mathcal C}} = \left(\begin{array}{cc}
{\rm cov}(A,A) & {\rm cov}(A,B) \\
{\rm cov}(B,A) & {\rm cov}(B,B) \end{array}\right)
\end{equation}
Then the eigenvalues of this matrix represent the semi-major and
semi-minor axes of the covariance ellipse, while the eigenvectors
provide the orientation of the ellipse.

\section{Results}
\label{Results}

Following the idea of building relations between values of model
parameters and macroscopic nuclear quantities~\cite{Chen:2010qx},
one finds that by changing the two solely isovector parameters,
which are $g_\rho$ and $\Lambda_{\rm v}$ in the RMF
model~\cite{Horowitz:2000xj}, and $x_0$ and $x_3$ in the SHF model,
only the symmetry energy $S(\rho)$ is modified while properties of
SNM such as saturation density $\rho_0$, binding energy per nucleon
at saturation density $E_0$, incompressibility coefficient at
saturation density $K_0$, and effective mass $M^{\ast}$ at
saturation all remain unchanged. Thus, in the following we optimize
the two isovector parameters [$F=2$ in Eqs.~(\ref{Taylor2}) and
(\ref{Covariance})] with respect to the available range of PNM EoSs
to constrain the values of $J$ and $L$ at saturation density by
employing the covariance analysis method discussed above.

\subsection{Reference models}
\label{Refmodels}

As representative RMF parameterizations, we choose the
accurately-calibrated NL3$^{\ast}$~\cite{Lalazissis:2009zz} and the
recent IU-FSU~\cite{Fattoyev:2010mx} parametrizations. The IU-FSU is
the recent parameterization that was validated against experimental,
observational, and theoretical data, while the accurately-calibrated
NL3$^{\ast}$ parameterization gives a much stiffer EoS of SNM
(larger value of $K_0$ and smaller value of $\zeta$ parameter) and a
stiff symmetry energy (larger values of symmetry energy $J$ and
slope $L$) and therefore offers a suitable contrast to IU-FSU.

To compare the RMF and SHF models on the same footing, we create two
Skyrme parameterizations which give the same properties of nuclear
matter at saturation as the two RMF parametrizations, herein
referred to as SkNL3$^{\ast}$ and SkIU-FSU forces, through the
method of writing the Skyrme parameters as functions of macroscopic
nuclear quantities~\cite{Chen:2009wv, Chen:2010qx}. Note that these two new
Skyrme parameterizations are intended only to serve as references in this study.

Several definitions of the nucleon effective mass exist in the
literature~\cite{vanDalen:2005ns}. In the RMF model the Dirac
effective mass is defined through the scalar part of the nucleon
self-energy in the Dirac equation:
\begin{equation}
M^{\ast}_{{\rm D}, q} = M_{q} + \Sigma^{\rm s}_{q} \ ,
\end{equation}
where the nucleon self-energy is given as $\Sigma^{\rm s}_{\rm n}
\equiv  \Sigma^{\rm s}_{\rm p}= -g_{\rm s} \phi$ in the RMF model
considered in this work. It has been well documented that there is a
strong correlation between the Dirac effective nucleon mass at
saturation density $M^{\ast}_{\rm D}$ and the strength of the
spin-orbit force in nuclei~\cite{Reinhard:1989zi, Gambhir:1989mp,
Bodmer:1991hz, Serot:1997xg}. Indeed, one of the most compelling
features of RMF models is the reproduction of the spin-orbit
splittings in finite nuclei. This occurs when the velocity
dependence of the equivalent central potential that leads to
saturation arises primarily due to a reduced nucleon effective
mass~\cite{Furnstahl:1997tk}. It is shown that models with effective
masses outside the range $0.58 < M^{\ast}_{\rm D}/M < 0.64$ will not
be able to reproduce empirical spin-orbit
couplings~\cite{Furnstahl:1997tk}, when no tensor couplings are
taken into account. On the other hand, the non-relativistic
effective mass parameterizes the momentum dependence of the single
particle potential, which is the result of a quadratic
parameterization of the single particle spectrum. A recent
study~\cite{Dutra:2012mb} puts a bound of $0.69 < M^{\ast}/M < 1.0$
for the non-relativistic effective masses. It has been
argued~\cite{Jaminon:1989wj} that the so-called Lorentz mass
$M^{\ast}_{\rm L}$  should be compared with the non-relativistic
effective mass extracted from analyses carried out in the framework
of nonrelativistic optical and shell models. For consistency, we
choose the effective mass in the Skyrme parameterizations to be
equal to the Lorenz mass in the RMF parameterizations. Since the RMF
model we use in this work gives the same isoscalar and isovector
effective masses, we set them equal in the reference SHF model too.

\begin{figure}[h]
\vspace{-0.05in}
\includegraphics[width=0.75\columnwidth,angle=0]{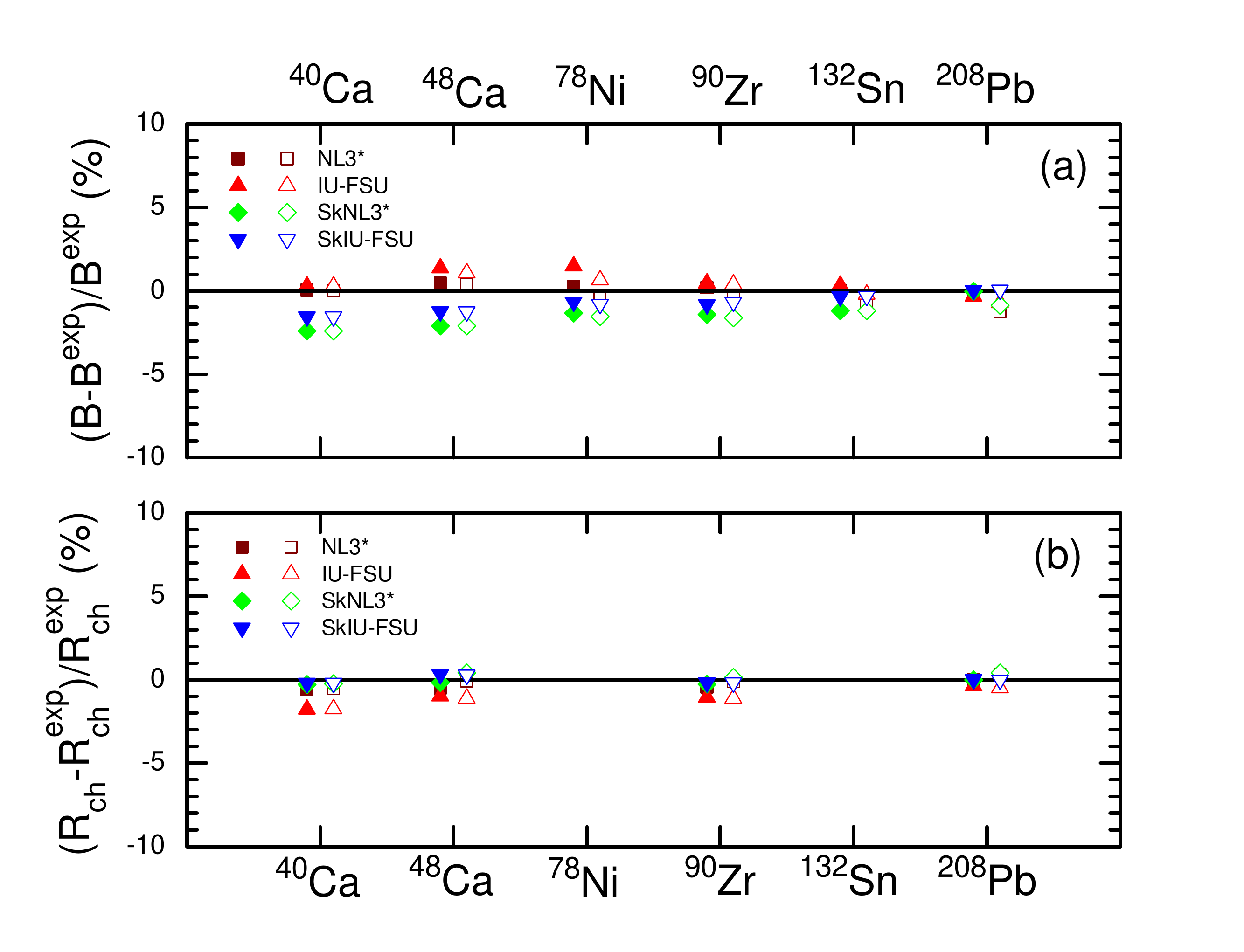}
\caption{(Color online) Relative deviation of the binding energies
(a) and charge radii (b) of closed-shell nuclei from the reference
models discussed in the text compared with the experimental data
(with superscript `exp') from Refs.~\cite{Audi:2002rp, Angeli:2004}.
Filled symbols are from the original parameterizations and empty
symbols are from the PNM modified parameterizations.} \label{Fig1}
\end{figure}

\begin{table}[h]
\begin{tabular}{|l||c|c|c|c|c|c|c|c|c|c|c|}
 \hline
 & $\rho_{0}~({\rm fm}^{-3}) $ & $E_{0}$~(MeV)
           & $K_{0}$~(MeV) & $M^{\ast}_{\rm D}$~($M$) & $M^{\ast}_{\rm L}$~($M$) & $M^{\ast}_{\rm S}$~($M$) & $M^{\ast}_{\rm V}$~($M$) & $J$~(MeV) & $L$~(MeV) & $K_{\rm sym}$~(MeV) & $R_{\rm skin}$~(fm) \\
\hline \hline
NL3$^{\ast}$     & 0.1500  & $-$16.32 & 258.49 & 0.594 & 0.671 & -     & -     & 38.7 & 122.7 & 105.7  & 0.29  \\
SkNL3$^{\ast}$   & 0.1527  & $-$15.76 & 258.49 & -     & -    & 0.671  & 0.671 & 38.7 & 122.7 & 62.7 & 0.27 \\
IU-FSU           & 0.1546  & $-$16.40 & 231.33 & 0.609 & 0.687 & -     & -     & 31.3 & 47.2  & 28.5 & 0.16 \\
SkIU-FSU         & 0.1575  & $-$15.70 & 231.33 & -     & -    & 0.687  & 0.687 & 31.3 & 47.2  & $-$132.0 & 0.16 \\
\hline
\end{tabular}
\caption{Macroscopic quantities from four reference
parameterizations. They are binding energy per nucleon $E_0$ and
incompressibility $K_0$ of SNM, Dirac ($M^{\ast}_{\rm D}$) and
Lorentz ($M^{\ast}_{\rm L}$) effective mass from the RMF model,
non-relativistic isoscalar ($M^{\ast}_{\rm S}$) and isovector
($M^{\ast}_{\rm V}$) effective mass from the SHF model, the symmetry
energy $J$, its slope parameter $L$ and curvature parameter $K_{\rm
sym}$ at saturation density, and the resulting neutron skin
thickness $R_{\rm skin}$ of $^{208}$Pb.} \label{Table1}
\end{table}

Finally, the isoscalar parameters of the two reference Skyrme forces
are then re-adjusted to fit the binding energy and charge radius of
$^{208}$Pb by adjusting only the saturation density $\rho_0$ and the
binding energy $E_0$ of SNM. As shown in Fig. \ref{Fig1}, these
models predict the charge radii and binding energies of other doubly
closed-shell nuclides within 1-2\% accuracy. We note that these
finite nuclei properties are obtained by solving the Dirac equation
for the RMF model and the Schr\"{o}dinger equation for the SHF
model. The bulk nuclear matter observables predicted by these
reference models are given in Table~\ref{Table1}. In terms of the
predicted values of isoscalar and isovector bulk observables, both
corresponding RMF and SHF models are therefore almost equivalent.

\begin{figure}[h]
\vspace{-0.05in}
\includegraphics[width=0.85\columnwidth,angle=0]{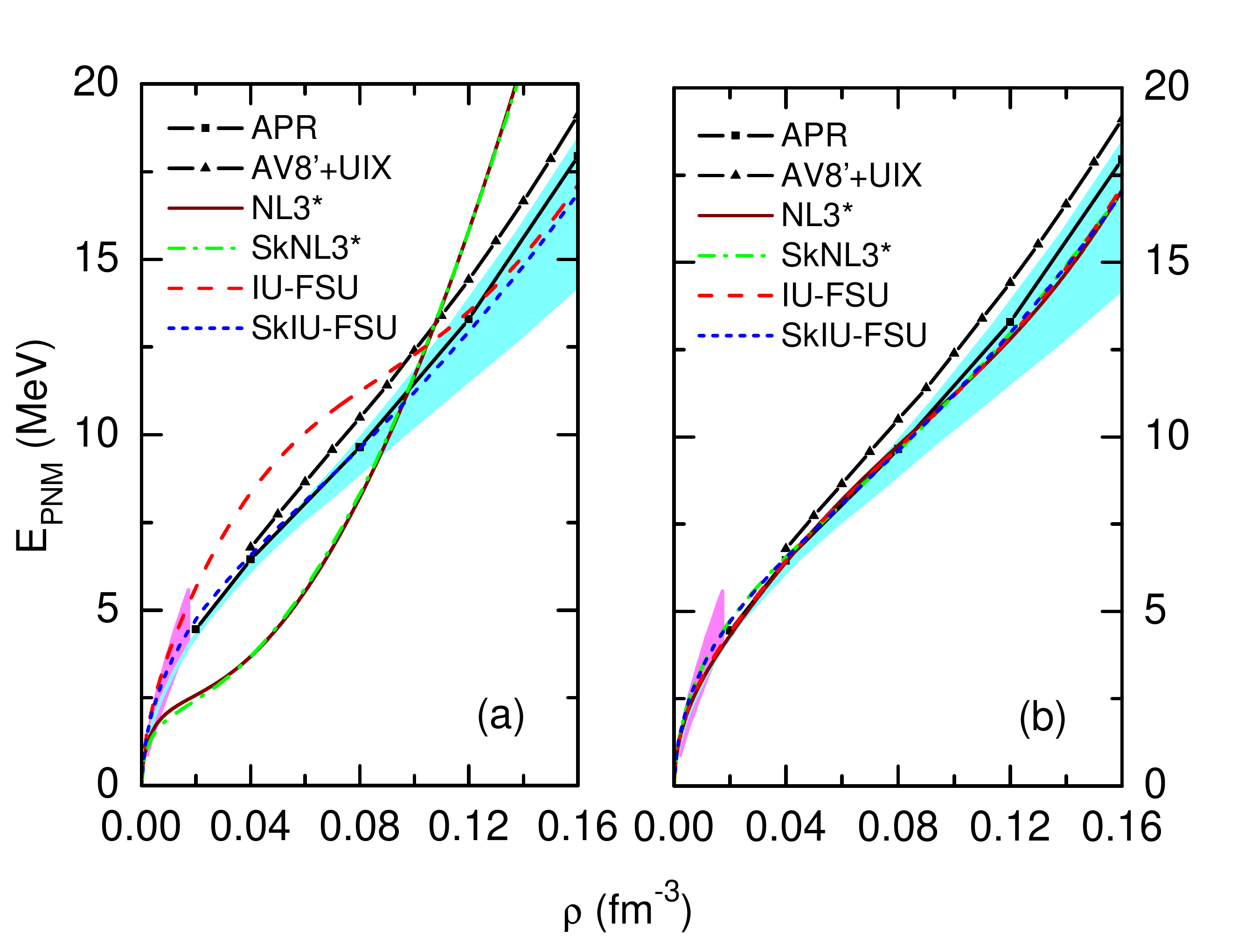}
\caption{(Color online) Comparing the PNM EoS from four reference
parameterizations with the AFDMC EoS in the AV8$^{\prime}$+UIX
Hamiltonian~\cite{Gandolfi:2011xu}, the variational APR EoS
~\cite{Akmal:1998cf}, the low-density band from the constraints of
resonant Fermi gases~\cite{Schwenk:2005ka}, and the high-density
band from the chiral effective field theory calculations with
3-neutron forces~\cite{Hebeler:2009iv}, before (a) and after (b) PNM
optimization.} \label{Fig2}
\end{figure}

The energy per neutron $E_{\rm PNM}$ predictions at sub-saturation
densities for our reference models are plotted on the left panel (a)
of Fig. \ref{Fig2}, compared to the results obtained by various
microscopic approaches. One can see that even among our four
parameterizations there is wide variance in the EoS of PNM at all
densities, and little agreement with those microscopic calculations.
The very wide range of predictions of the symmetry energy parameters
and the corresponding widespread predictions for the neutron skins
of nuclei inherent in these parameterizations are seen in Table
\ref{Table1}.

\begin{table}
\begin{tabular}{|l||c|c|c|c|c|c|c|c|c|c|c|c|}
 \hline
          & $S_{0.1}^0$ & $S_{0.1}$ & $J^0$ & $J$           & $L^0$ & $L$            & $K_{\rm sym}^0$ & $K_{\rm sym}$ &  $K_{\tau}^0$  &  $K_{\tau}$  &  $R_{\rm skin}^0$  &  $R_{\rm skin}$ \\
\hline \hline
NL3$^{\ast}$     & 25.7 & 24.9 $\pm$ 0.4 & 38.7 & 30.7 $\pm$ 0.7 & 122.7 & 50.3 $\pm$ 1.8 & 105.7 &\phantom{xx}39.2 $\pm$ 17.8  & -684.4 & -284.6 $\pm$ 29.4 & 0.29 & 0.18 $\pm$ 0.01\\
SkNL3$^{\ast}$   & 25.0 & 24.5 $\pm$ 0.3 & 38.7 & 31.0 $\pm$ 0.9 & 122.7 & 46.4 $\pm$ 6.4 & \phantom{x}62.7 &-156.0 $\pm$ 16.6  & -529.4 & -380.0 $\pm$ 15.2 & 0.27 & 0.16 $\pm$ 0.01\\
IU-FSU           & 25.7 & 24.9 $\pm$ 0.4 & 31.3 & 31.4 $\pm$ 0.7 & \phantom{x}47.2 & 52.9 $\pm$ 2.0 & \phantom{x}28.5 &\phantom{xx}-6.8 $\pm$ 12.9  & -195.3 & -257.6 $\pm$ 22.3 & 0.16 & 0.18 $\pm$ 0.01 \\
SkIU-FSU         & 24.4 & 24.4 $\pm$ 0.3 & 31.3 & 31.4 $\pm$ 0.9 & \phantom{x}47.2 & 48.0 $\pm$ 6.2 & -132.0 &-130.2 $\pm$ 13.3  & -343.9 & -345.6 $\pm$ 15.3 & 0.16 & 0.16 $\pm$ 0.01 \\
\hline
\end{tabular}
\caption{Isovector observables and associated 1$\sigma$ error bars
from four reference parameterizations before (with superscript `0')
and after (without superscript `0') the PNM constraints are applied.
Values are shown for the symmetry energy at $\rho = 0.1$ ${\rm fm}^{-3}$ $S_{0.1}$
and at saturation density $J$, slope parameter $L$, curvature
parameter $K_{\rm sym}$, isospin-dependent part of incompressibility
$K_{\tau}$, and the neutron skin thickness $R_{\rm skin}$ of
$^{208}$Pb. All the quantities are in MeV apart from $R_{\rm skin}$
which is in fm.} \label{Table2}
\end{table}

\subsection{Symmetry Energy Coefficients}
\label{Results1}

Having established our reference models, we next minimize the
$\chi^2$ with respect to the PNM constraints~\cite{Akmal:1998cf,
Hebeler:2009iv, Gandolfi:2011xu} in the density range of $0.04 \leq
\rho \leq 0.16$ fm$^{-3}$ by adjusting two isolated (solely
isovector) parameters. This leads to optimized values of the
model parameters and thus the density dependence of symmetry energy up to
saturation density once the EoS of SNM is fixed. All isoscalar
parameters remain unchanged and there is very little change in the
prediction of binding energies and charge radii as is shown in Fig. \ref{Fig1}.

As can be seen in panel (b) of Fig. \ref{Fig2}, we obtain the EoS of
PNM for a given RMF or SHF parameterization that best fits within
the band of microscopic PNM calculations at the minimum of the
$\chi^2$-function. The resulting RMF and SHF models predict very
similar symmetry energies $J$, while the RMF model predicts a
consistently higher central value for $L$ by about 4-5 MeV than the
SHF model as shown in Table \ref{Table2}.

\begin{figure}[h]
\vspace{-0.05in}
\includegraphics[width=0.85\columnwidth,angle=0]{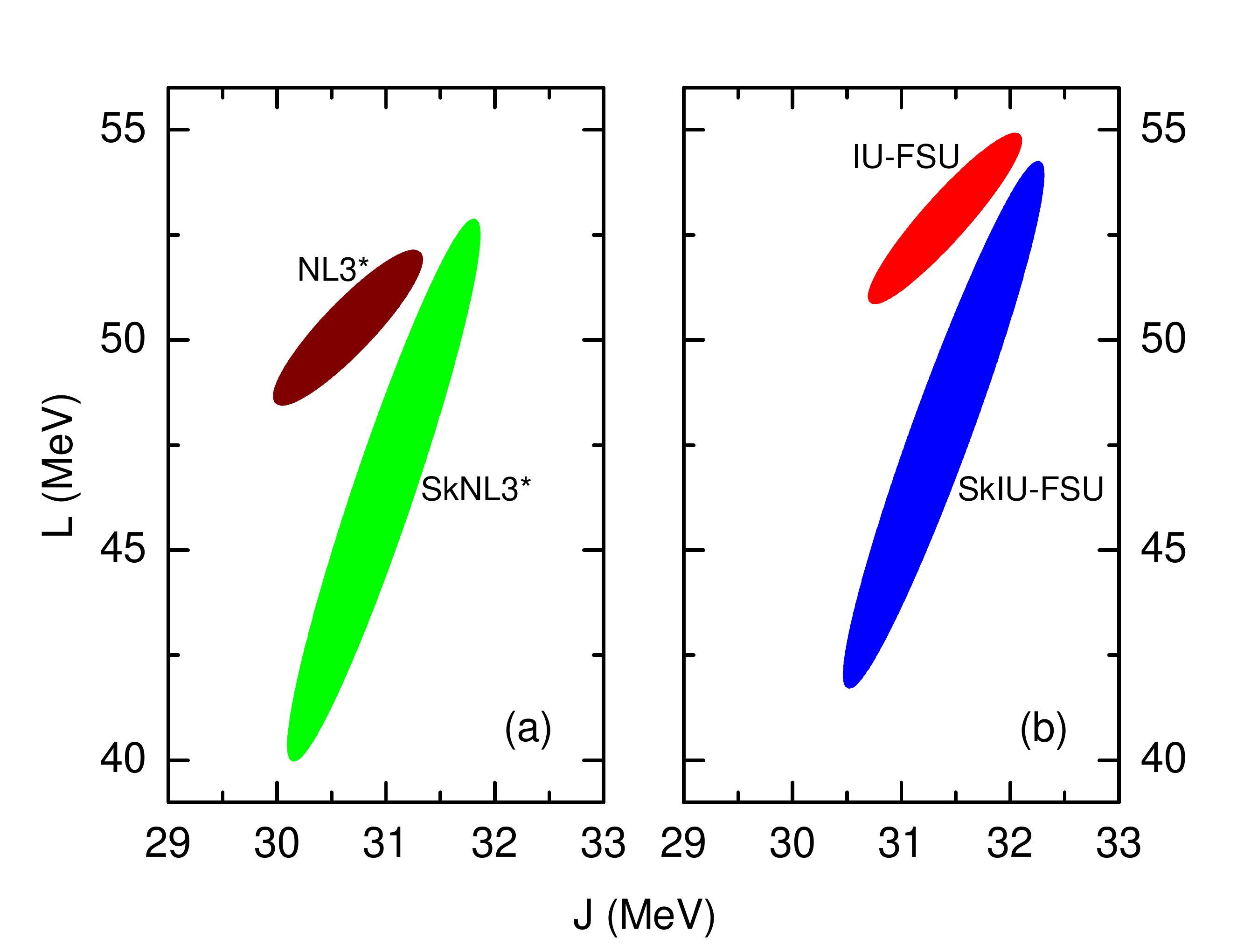}
\caption{(Color online) $1\sigma$ joint confidence regions for the
symmetry energy $J$ and its slope parameter $L$ at saturation
density for the RMF and SHF models.} \label{Fig3}
\end{figure}

The 1$\sigma$ errors on these two isolated parameters can be
translated into equivalent errors on $J$ and $L$. The errors in $J$
are less than $\pm 1$ MeV for all the parameterizations. The RMF
model gives a relatively small error in $L$ of around $\pm 2$ MeV,
while the SHF model gives a much larger error around $\pm 6$ MeV.
Table \ref{Table2} appears to indicate that within the $1\sigma$
errors, both models are consistent in their predicted values of $J$
and $L$. However, in Fig. \ref{Fig3} we plot a $1\sigma$ joint
confidence regions in the $J$-$L$ plane for both RMF and SHF models,
thus showing that in fact the two models predict non-overlapping
regions in $J$-$L$ space. Both models show a positive correlation
between $J$ and $L$, but with differing slopes. For example, for
IU-FSU and SkIU-FSU parametrizations the relations are approximately
\begin{eqnarray}
L &=&  \left(2.4 \, J - 23\right) \, {\rm MeV}, \qquad {\rm (RMF)} \  \notag \\
L &=&  \left(6.0 \, J - 140\right) \, {\rm~MeV}, \qquad {\rm (SHF)}
\
\end{eqnarray}
within the constraints of $J$ and $L$ shown in Table \ref{Table2}.

\begin{figure}[h]
\vspace{-0.05in}
\includegraphics[width=0.85\columnwidth,angle=0]{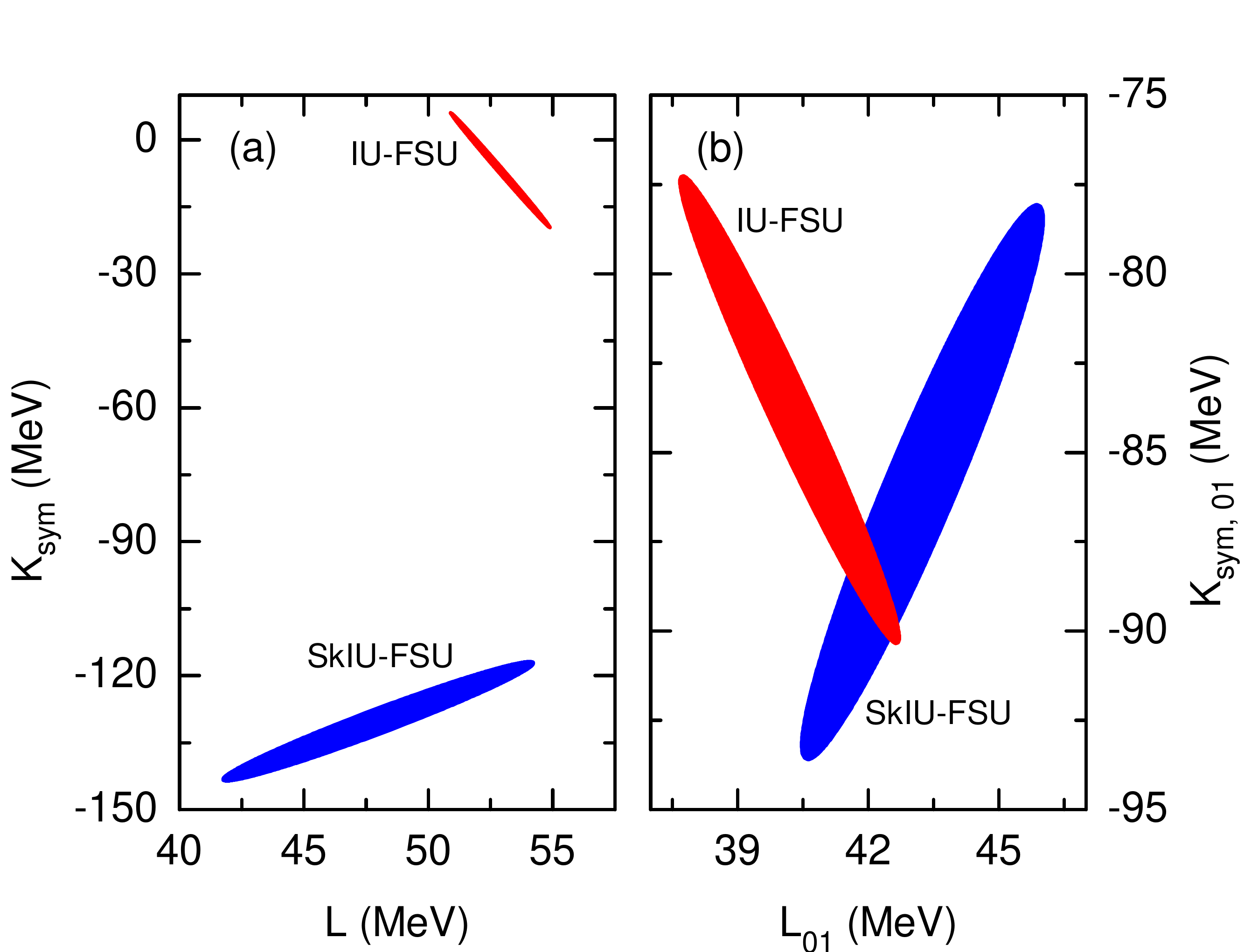}
\caption{(Color online) $1\sigma$ joint confidence regions for the
slope parameter $L$ and curvature parameter $K_{\rm sym}$ of the
symmetry energy at saturation density (a) and at $\rho =0.1$
fm$^{-3}$ (b) from the IU-FSU and SkIU-FSU parameterizations.}
\label{Fig4}
\end{figure}

The origin of this difference lies mainly in the values of the
higher-order symmetry energy parameters that are predicted upon
optimization. There is a strong model dependency in the prediction
for the curvature parameter of the symmetry energy $K_{\rm sym}$
(see Table \ref{Table2}). For example, after the PNM optimization
IU-FSU predicts $K_{\rm sym} = -6.8 \pm 12.9$ MeV, while its
Skyrme-like version predicts a smaller value of $K_{\rm sym} =
-130.2 \pm 13.3$ MeV. When we plot the $1\sigma$ joint confidence
regions in the $K_{\rm sym}$-$L$ plane for both RMF and SHF models
(see the left panel (a) of Fig. \ref{Fig4}) further differences can
be seen: there is, generically, a negative correlation between the
slope of the symmetry energy and $K_{\rm sym}$ in the RMF model,
while this correlation is positive in the case of the SHF model.
Only at a sub-saturation density of $\rho = 0.1$ fm$^{-3}$ do the
two models have similar values of $K_{\rm sym} \left(\rho=0.1 \,
{\rm fm}^{-3}\right)$ (see the right panel (b) of Fig. \ref{Fig4}),
although the correlations are still opposite. We emphasize that
these qualitative features emerge whatever the starting
parameterization of the RMF or SHF model used.

It is widely accepted that the Giant Monopole Resonance (GMR)
provides the cleanest and most direct route to the nuclear
incompressibility around normal density~\cite{Piekarewicz:2008nh}.
It has been also proposed that GMR energies of finite nuclei as well
as the nuclear matter incompressibility should be computed within
the same theoretical framework~\cite{Blaizot:1980tw, Blaizot:1995}.
The expression for the incompressibility of neutron-rich matter at
saturation density is given by~\cite{Piekarewicz:2008nh}:
\begin{equation}
K_{\rm sat}(\alpha) = K_0 + K_{\tau} \alpha^2 +
\mathcal{O}(\alpha^4) \ ,
\end{equation}
where the coefficient of $\alpha^2$ is
\begin{equation}
K_{\tau} = K_{\rm sym} - 6L - \frac{Q_0}{K_0} L \
\end{equation}
with $Q_0$ being the skewness of SNM~\cite{Chen:2009wv}. Although
both RMF and SHF models used in this work share the same value of
$K_0$, their predictions of $K_{\rm sat}$ are different due to the
difference in $K_{\tau}$, which in turn is mainly due to the
difference in $K_{\rm sym}$. In Table \ref{Table3} we provide the
values of $K_{\rm sat}$ for different values of isospin asymmetry.
Due to the small values of isospin asymmetry in finite nuclei, the
difference of the incompressibility for different models is in fact
small. Comparing with the constraint of $-760 < K_{\tau} < -372$ MeV
in Ref.~\cite{Dutra:2012mb} extracted directly from the GMR data,
both RMF and SHF models predict marginally consistent or slightly
higher values of $K_{\tau}$ after the PNM optimization as shown in
Table~\ref{Table2}, suggesting that both RMF and SHF models have
difficulty in simultaneously predicting GMR properties consistent
with experiment and the PNM EoS consistent with our best theoretical
calculations.

  \begin{table}[h]
  \begin{center}
  \begin{tabular}{|c|c|c|c|}
    \hline
     & $K_{\rm sat}(\alpha = 0)$ (MeV) & $K_{\rm sat}(\alpha = 0.111)$ (MeV) & $K_{\rm sat}(\alpha = 0.212)$ (MeV) \\
    \hline
    \hline
    NL3$^{\ast}$    & $258.5$ & $255.0$ & $245.8$  \\
    SkNL3$^{\ast}$  & $258.5$ & $253.8$ & $241.5$  \\
    IU-FSU          & $231.3$ & $228.2$ & $219.8$ \\
    SkIU-FSU        & $231.3$ & $227.1$ & $215.9$ \\
    \hline
  \end{tabular}
\caption{Incompressibility of neutron-rich matter with different
isospin asymmetries $\alpha=0$ (SNM), $0.111$ (${}^{90}$Zr), and
0.212 (${}^{208}$Pb) at saturation density from the four
parameterizations after the PNM optimization.}
  \label{Table3}
  \end{center}
 \end{table}

Different values of the bulk properties of SNM will affect the PNM
constraints on the symmetry energy. For example, the saturation
density $\rho_0$, the binding energy at saturation $E_0$, and the
incompressibility coefficient at saturation $K_0$ will affect the
EoS of SNM and thus modify slightly the optimized symmetry energy
from a fixed set of PNM EoS constraints. The effective mass
$M^{\ast}$ dominates these uncertainties in the results of the PNM
optimization. As can be seen from the expression for the symmetry
energy in the RMF model~\cite{Horowitz:2001ya}:
\begin{equation}
\label{SymmetryEnergyRMF}
  S(\rho) = \frac{k^{2}_{F}}{6 \sqrt{k_F^2 + M^{\ast 2}}} +
  \frac{g_{\rho}^{2}\rho}{8m_{\rho}^{\ast 2}}\;, \quad
  \Big(m_{\rho}^{\ast 2} \equiv m_{\rho}^{2}+
   2\Lambda_{\rm v}g_{\rho}^{2}\left(g_{\rm v}V_{0}\right)^{2}\Big)
   \;
\end{equation}
$M^{\ast}$ affects the kinetic contribution to the symmetry energy
while adjusting $g_\rho$ and $\Lambda_{\rm v}$ only modifies the
potential contribution to the symmetry energy. We find that
increasing the effective mass at saturation by $\sim 10\%$ decreases
the optimized value of the slope of the symmetry energy at
saturation density $L$ by $\sim 10$ MeV. The isovector effective
mass, here set equal to the isoscalar effective mass in the SHF
model to be consistent with the RMF models, affects the value of $L$
obtained in the PNM optimization by the same order of magnitude, but
in the opposite direction. The curvature of the symmetry energy
$K_{\rm sym}$ is changed by a much smaller relative amount.
Therefore the 1-$\sigma$ confidence ellipses change their positions
in the $J$-$L$ plane as the SNM  properties are varied, but they
retain very similar values of their slopes, and the RMF and SHF
confidence ellipses maintain their relative positions. Similarly,
the $K_{\rm sym}$-$L$ confidence ellipses change their $L$-position
upon variation of SNM properties, but retain their $K_{\rm sym}$
values and relative orientation and spacing.

In order to get a better idea of the overall range of predictions
for $J$, $L$ and $K_{\tau}$ taking the additional model parameters
into account, we take 11 RMF parameterizations and 73 SHF
parameterizations from the literature that have been created since
1995 \cite{Chen:2007ih,Dutra:2012mb}. We optimize the pure isovector
parameters of each parameterization to the PNM results and examine
the resultant constraints; these are displayed in Table
\ref{Table4}.

\begin{table}[t]
\begin{tabular}{|l||c|c|c|}
 \hline
 & $J$~(MeV) & $L$~(MeV) & $K_{\tau}$~(MeV) \\
\hline \hline
RMF   & 30.2  --- 31.4 & 36.1 --- 59.3 & -329.7 --- -215.7  \\
SHF  & 30.1 --- 33.2 & 28.5 --- 64.4 & -418.8 --- -235.3   \\
\hline
\end{tabular}
\caption{Predicted ranges for symmetry energy parameters within RMF
and SHF models with their pure isovector parameters optimized to PNM
and taking into account all remaining variation from
parameterizations constructed since 1995.} \label{Table4}
\end{table}

\subsection{Implications for predictions of neutron skin thicknesses and neutron star radii}
\label{Results2}

Measurements of the neutron skin thicknesses of various nuclides
using strong interaction probes~\cite{Ray:1978ws, Ray:1979qv,
GarciaRecio:1991wk, Starodubsky:1994xt,
Clark:2002se,Trzcinska:2001sy, Krasznahorkay:1994,
Krasznahorkay:1999zz, Klimkiewicz:2007zz, Terashima:2008zza} and,
recently, weak interaction probes~\cite{Abrahamyan:2012gp,
Ban:2010wx} in the PREX experiment, are an important tool to probe
the density dependence of the symmetry energy due to the very close
correlation of $L$ to the size of the neutron skin in neutron-rich
nuclides  \cite{Brown:2000, Furnstahl:2001un, Centelles:2008vu,
Centelles:2010qh, Vidana:2009is}. Since our optimized RMF and SHF
models give nearly matching ranges of $L$, we expect the neutron
skin predictions to be similar.

\begin{figure}[h]
\vspace{-0.05in}
\includegraphics[width=0.75\columnwidth,angle=0]{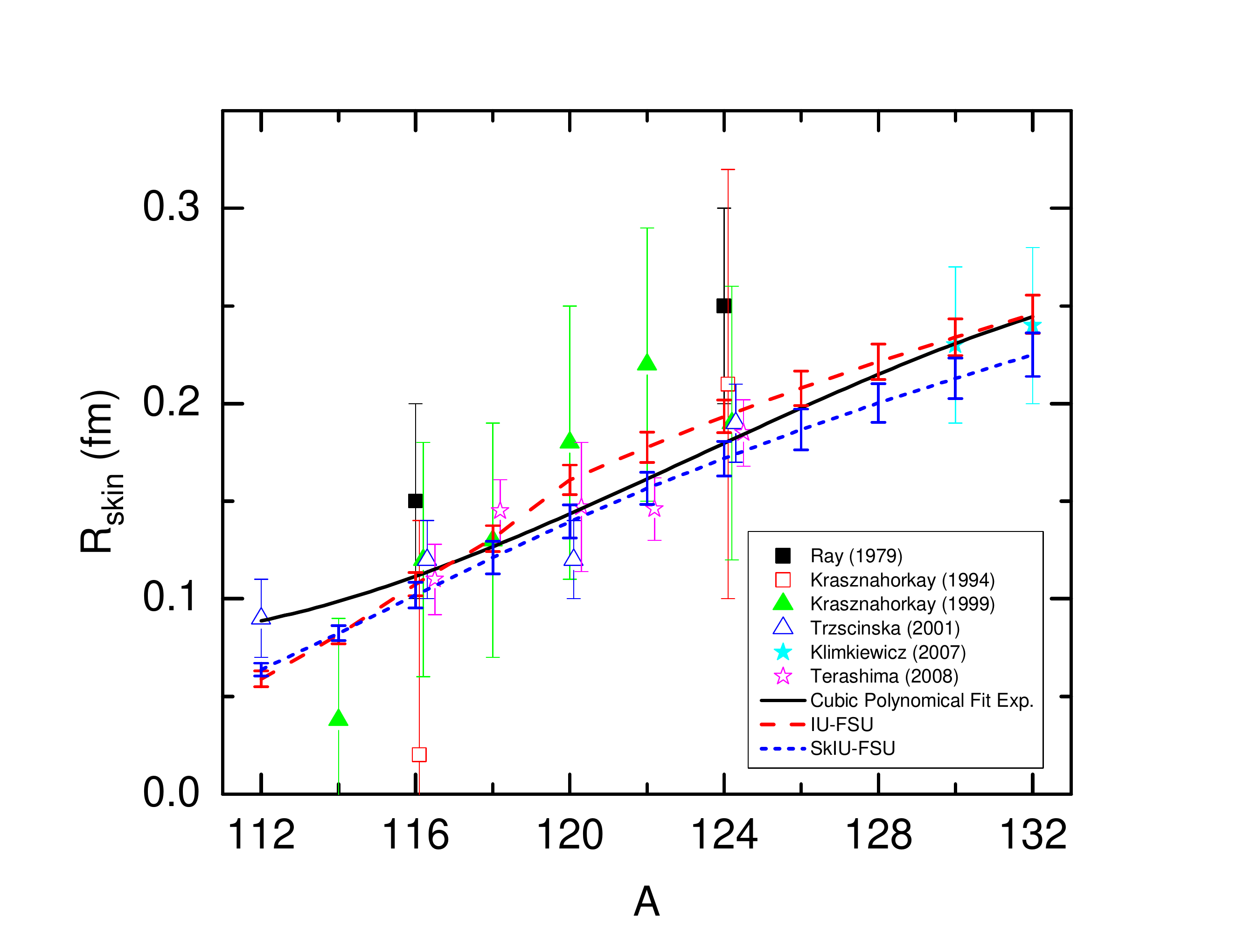}
\caption{(Color online) Comparing the predictions of the neutron
skin thickness for Sn isotopes  from the IU-FSU and SkIU-FSU models
after the PNM optimization with those from different experimental
methods.} \label{Fig5}
\end{figure}

In Fig. \ref{Fig5}, we compare predictions of neutron skin
thicknesses from the IU-FSU and SkIU-FSU parameterizations to the
currently existing data on the neutron skin thickness of Tin
isotopes~\cite{Ray:1979qv, Krasznahorkay:1994, Krasznahorkay:1999zz,
Trzcinska:2001sy, Klimkiewicz:2007zz, Terashima:2008zza}. As
expected, both the post-optimization IU-FSU and SkIU-FSU models
agree well with the experimental data, with the RMF model giving a
systematically slightly higher value than the SHF model in all but
the lightest isotopes calculated. Thus consistency with our best
knowledge of the PNM EoS can be achieved simultaneously with
consistency of neutron skin predictions with current experimental
data within the RMF and SHF models.

The IU-FSU parameterization predicts $R_{\rm skin} = 0.18 \pm 0.01$
fm for $^{208}$Pb, while SkIU-FSU predicts a slightly lower value of
$R_{\rm skin} = 0.16 \pm 0.01$ fm (Table \ref{Table2}). The smaller
value of $R_{\rm skin}$ for SkIU-FSU is primarily due to model
dependence, which leads to a smaller value of optimized $L$ from the
PNM constraints. The current PREX obtained value for the neutron
skin thickness of lead is $R_{\rm skin} = 0.33^{+0.16}_{-0.18}$
fm~\cite{Abrahamyan:2012gp}. If the new PREX experiment reduces the
error bars without moving the central value for the neutron skin,
almost all current models of the nuclear structure would need to be
modified. Also, this would appear to call for a significant
modification of the PNM microscopic calculations.

Finally, we examine how the different symmetry energy
characteristics of RMF and SHF models are manifest in neutron star
radius predictions. Using our four post-optimization
parameterizations, we apply the EoS of $\beta$-stable and charge
neutral neutron star matter composed of neutrons, protons,
electrons, and muons throughout the core of the star. For the very
low density outer crust we use the BPS equation of
state~\cite{Baym:1971pw}. The equation of state of the inner crust
is approximated by the polytropic equation of state of the form $P =
A\mathcal{E}^{4/3}+B$~\cite{Carriere:2002bx}, where $A$ and $B$ are
determined to match the EoS at the boundaries of the inner crust.
Using our equations of state, we integrate the general relativistic
equation for hydrostatic equilibrium (the Tolman-Oppenheimer-Volkoff
equation) from the center to the surface of the star.

\begin{figure}[h]
\vspace{-0.05in}
\includegraphics[width=0.85\columnwidth,angle=0]{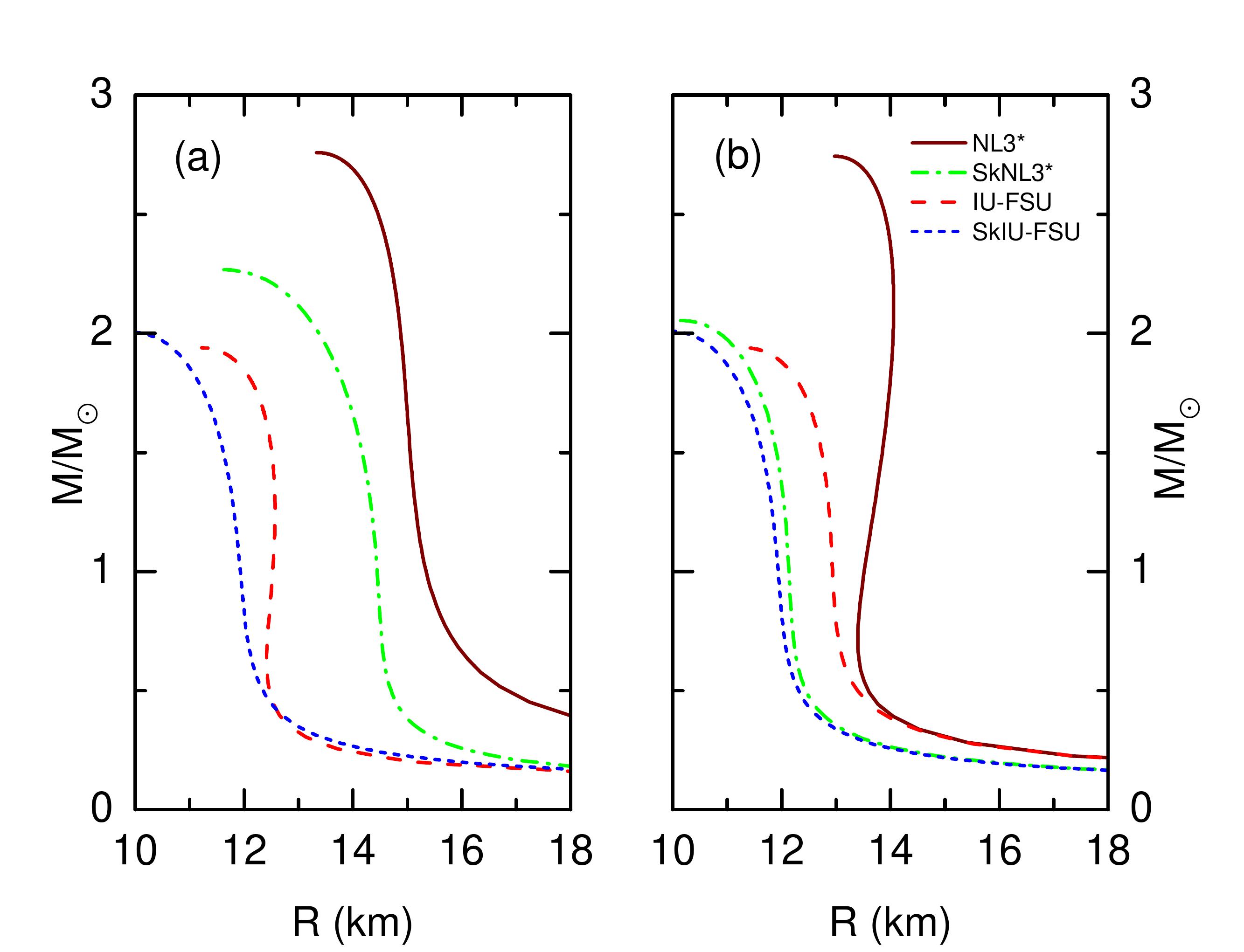}
\caption{(Color online) Mass-Radius relation of neutron stars
calculated from the four parameterizations before (a) and after (b)
the PNM optimization. } \label{Fig6}
\end{figure}

The reference RMF and SHF parameterizations before the PNM
optimization predict a wide range of results for low mass neutron
star radii as shown in the left panel (a) of Fig. \ref{Fig6}. In
particular, for a $1.0$ solar mass neutron star the difference in
the predictions of radii for the NL3$^{\ast}$ and IU-FSU equations
of state is equal to $\Delta R_{1.0} \approx 2.8$ km. There is a
similar difference between the original SkNL3$^{\ast}$ and SkIU-FSU
equation of state predictions, i.e., $\Delta R_{1.0} \approx 2.5$
km. This can be mainly attributed to the density dependence of the
symmetry energy, which is quite different in the two
parameterizations. Once calibrated to the PNM results, this
difference almost vanishes within the same model as shown in the
right panel (b) of Fig. \ref{Fig6}, i.e., both RMF and SHF
parameterizations now match each other more closely (excepting the
differences at high masses between the RMF models, a result of a
stiffer EoS of SNM in NL3$^{\ast}$ parameterizations at several
times saturation density due to the $\zeta$ parameter). Although
both NL3$^{\ast}$ and IU-FSU parameterizations in a given RMF or SHF
model predict similar radii, there is a clear difference between the
RMF and the SHF predictions as a whole. In the case of IU-FSU and
SkIU-FSU we have almost a $\sim 1$ km difference for the radius of a
canonical neutron star. This discrepancy is even larger in the case
of NL3$^{\ast}$, which is about $\sim 1.8$ km. Thus, there is a
strong model dependence when the two models are applied to neutron
star structure calculations after the same PNM optimization.

\begin{figure}[h]
\vspace{-0.05in}
\includegraphics[width=0.75\columnwidth,angle=0]{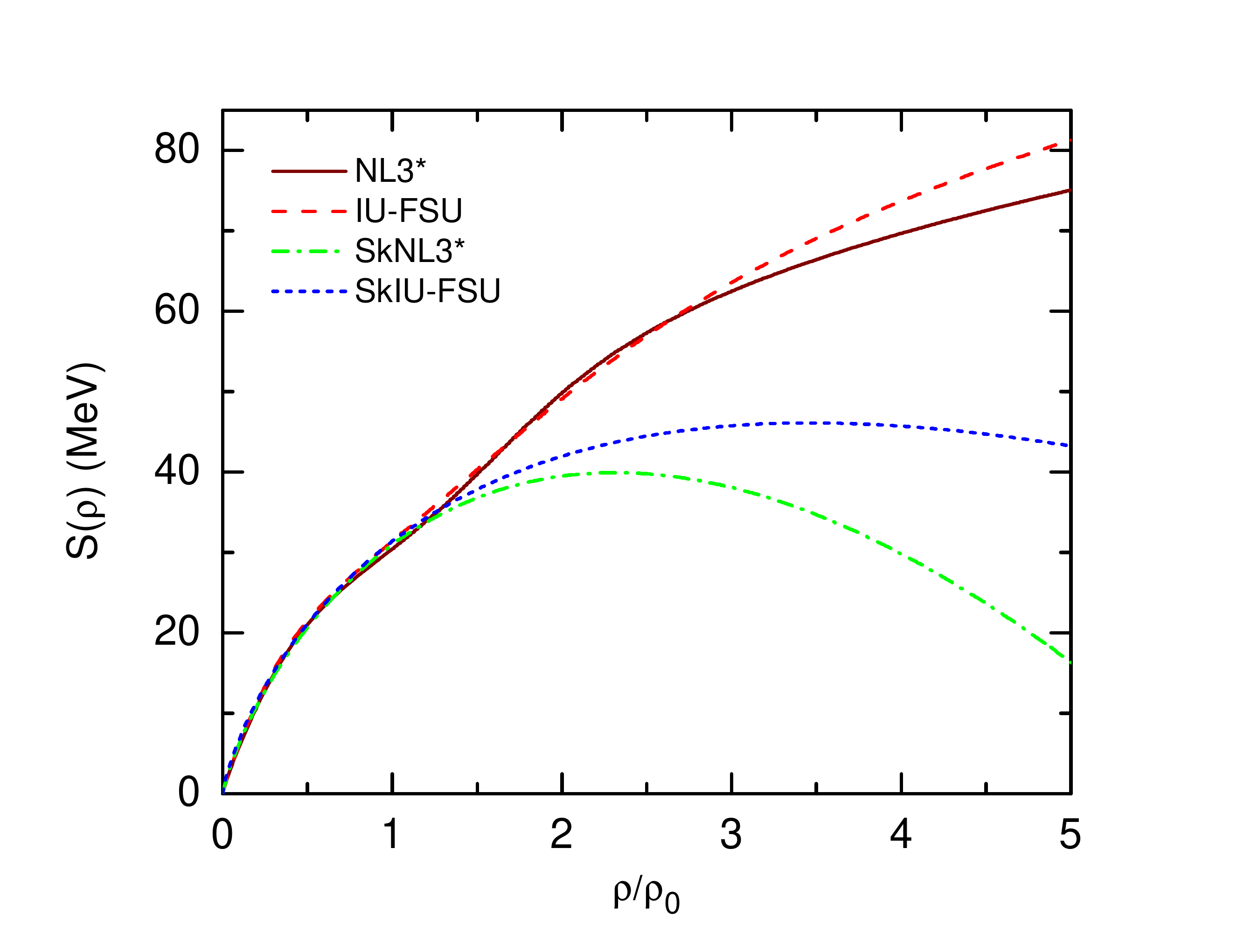}
\caption{(Color online) Density dependence of symmetry energy from
the four parameterizations after the PNM optimization.} \label{Fig7}
\end{figure}

The above model dependence actually comes from different density
dependence of symmetry energy at supra-saturation densities, flagged
by the model dependent difference in predictions of the curvature of
the symmetry energy at saturation density $K_{\rm sym}$. In Fig.
\ref{Fig7} we plot the density dependence of the symmetry energy for
the RMF and SHF models under consideration after the PNM
optimization. Note that the symmetry energy is almost the same in
all the models up to $\sim 1.5\rho_0$ saturation density. However,
the symmetry energy in the RMF functional is a monotonic increasing
function of density, while the SHF functional tends to give a
decreasing symmetry energy with increasing density at higher
densities. Again, this property is generic once the model has been
optimized to PNM EoS. The reason for this difference is manifest in
the functional forms of the symmetry energy given as:
\begin{eqnarray}
&& \label{SymEnerRMF} S_{\rm RMF}(\rho) = A(\rho) \rho^{2/3} + B(\rho) \rho \ , \\
&& \label{SymEnerSHF} S_{\rm SHF}(\rho) = a \rho^{2/3} - b \rho - c
\rho^{5/3} - d \rho^{\sigma + 1} \ ,
\end{eqnarray}
where $A(\rho)$ and $B(\rho)$ are positive-valued functions of
density [see Eq. (\ref{SymmetryEnergyRMF})], $a \equiv
\frac{\hbar^2}{6M}\left(\frac{3\pi^2}{2}\right)^{2/3}$ and $b$, $c$,
$d$ are constants that depend on Skyrme parameters only. The
symmetry energy in the RMF model is always positive as given in Eq.
(\ref{SymEnerRMF}), while certain terms of the symmetry energy in the
SHF model can become negative at higher densities [see Eq.
(\ref{SymEnerSHF})].

Recently, it was shown that currently available neutron star mass
and radius measurements provide significant constraints on the EoS
of PNM all the way up to several saturation
densities~\cite{Steiner:2011ft}. While this is true, we also show
that the low-density PNM constraints alone result in a pronounced
model dependency of radius predictions, as different masses and
radii can be obtained with the similar saturation properties
constrained by the low-density PNM EoS. Although our PNM
optimization tightly constrains the symmetry energy up to a little
above the saturation density, in order to understand its behavior at
higher densities, which is also important in determining neutron
star radii, one must rely on the heavy-ion collision
experiments~\cite{Danielewicz:2002pu,Li:2008gp,Baran:2004ih} and
neutron star observations~\cite{Lattimer:2006xb, Steiner:2004fi}.

\section{Conclusions}
\label{summary}

Using parameterizations of RMF and SHF energy-density functionals
prepared to give equally good fits to ground state properties of
doubly magic nuclei and identical symmetric nuclear matter
properties, and are fit to state-of-the-art {\it ab initio}
theoretical calculations of PNM up to saturation density, we have
conducted a systematic examination of the resultant predictions from
both models of the symmetry energy as a function of density and some
important terrestrial nuclear and neutron star observables sensitive
to $S(\rho)$.

We show that such RMF and SHF models result in very similar
predictions for the symmetry energy $J$ and its slope parameter $L$
at saturation density from both models so long as the isoscalar
effective mass from the SHF model is chosen to be equal to the
Lorenz effective mass from the RMF model, which is tightly
constrained around $\approx 0.7M$. Both models then give $J \approx
31.0 \pm 1$ MeV. The SHF parameterizations give values around $46$
--- $49$ $\pm 6$ MeV and the RMF parameterizations $50$ --- $53$ $\pm 2$
MeV for $L$. Resulting predictions of neutron skin thicknesses
$R_{\rm skin}$ for Sn isotopes and $^{208}$Pb therefore agree
closely and are consistent with the available experimental data.

When the 1$\sigma$ error bounds are plotted as ellipses in the
$J$-$L$ plane, a positively-correlated relationship between $J$ and
$L$ is observed for both models. However, different slopes are
obtained from the RMF and SHF models, and the two ellipses have no
overlapping area in the plane. This model dependence comes from the
different values of $K_{\rm sym}$ and higher-order symmetry energy
parameters; i.e. from the different functional form of the symmetry
energy implicit in the models.  Although the PNM constraints lead to
broadly similar behaviors of the symmetry energy as a function of
density up to $\approx 1.5 \rho_0$, they deviate significantly at
higher densities due to the differences in the functional form of
the symmetry energy. With the same PNM constraints up to the
saturation density, the RHF model tends to predict a rising symmetry
energy at higher densities, whereas the SHF model predicts a
symmetry energy that may decrease with density at higher densities,
and thus leading to the uncertainty of up to $\sim 2$ km in neutron
star radii. Care must therefore be taken in extracting constraints
on the symmetry energy, particularly on $J$ and $L$, from inferred
neutron star radii within one particular model.

The absolute values of the predictions are found to be mainly
sensitive to the effective mass, with increases (decreases) of $\sim
0.1$ $M$ leading to decreases (increases) of $L$ by $\sim 10$ MeV.
We confirm this systematic analysis by analyzing the predictions
from 11 RMF and 73 SHF parameterizations constructed since 1995,
finding overall ranges taking into account remaining freedom in the
parameter values, of of $30 \lesssim J \lesssim 31.5$ MeV, $35
\lesssim L \lesssim 60$ MeV, $-330 \lesssim K_{\tau} \lesssim -216$
MeV for RMF models  and $30 \lesssim J \lesssim 33$ MeV, $28
\lesssim L \lesssim 65$ MeV, $-420 \lesssim K_{\tau} \lesssim -325$
for SHF models.

Notably, some recent constraints inferred from experimental data on
giant monopole resonances of Sn and Cd isotopes
\cite{Li:2007bp,Sagawa:2007sp,Garg:2011yr} and on neutron skins
\cite{Centelles:2008vu} place $K_{\tau}$ in the overall range $-650
< K_{\tau} < -375$ MeV. It has been pointed out that these results
are inconsistent with many individual Skyrme parameterizations and
microscopic nuclear matter calculations
\cite{Pearson:2010zz,Dutra:2012mb}; our results generalize these
points to demonstrate that these particular $K_{\tau}$ constraints
are inconsistent with the RMF model as a whole and only marginally
consistent with SHF models as a whole, within 1$\sigma$ confidence
intervals resulting from optimization to PNM calculations. Thus,
either the density dependence of RMF and SHF models is insufficient
to simultaneously describe PNM within current bounds and GMR/neutron
skin experimental data, or there are overlooked problems with the
extraction of the $K_{\tau}$ constraints in the above works and the
error bounds are underestimated, as has been suggested
\cite{Pearson:2010zz}. Note that the $K_{\tau}$ ranges we extract
from both models are consistent with another $K_{\tau}$ constraint
extracted from isospin diffusion in heavy ion collisions $-490 <
K_{\tau} < - 250$ MeV \cite{Chen:2009wv}. More work needs to be done
to check these hypotheses while taking the dependency of functional
forms (e.g. \cite{Erler:2010zh}) into consideration.

\begin{acknowledgments}
The authors are grateful to Prof. Lie-Wen Chen for providing the
experimental data of neutron skin thickness of Tin isotopes, to Dr.
Stefano Gandolfi for making available the AFDMC data, and to Prof.
Jorge Piekarewicz for providing the Hartree code to calculate the
ground state properties of finite nuclei in the RMF model, and for
many fruitful discussions. This work is supported in part by the
National Aeronautics and Space Administration under grant NNX11AC41G
issued through the Science Mission Directorate, and the National
Science Foundation under Grants No. PHY-1068022 and No. PHY-0757839.
\end{acknowledgments}

\bibliography{ReferencesFJF}

\end{document}